\documentclass[10pt,twocolumn,letterpaper]{article}
\usepackage[left=54pt,right=54pt,top=84pt,textheight=626pt,footskip=24pt]{geometry}
\usepackage[T1]{fontenc}
\usepackage[utf8]{inputenc}
\usepackage[tt=false,type1=true]{libertine}
\usepackage[varqu]{zi4}
\usepackage{amsmath,amssymb}
\usepackage[libertine]{newtxmath}
\usepackage{booktabs,array,graphicx}
\usepackage{microtype}
\usepackage[font=small,labelfont=bf]{caption}
\usepackage[numbers,sort&compress]{natbib}
\usepackage{xurl}

\usepackage[hidelinks]{hyperref}
\hypersetup{pdftitle={M3FT: Micro--Macro--Micro Reasoning for Multi-Agent Cryptocurrency Perpetual Futures Trading},pdfauthor={Jiashun Wang, Beile Wang, Guangyan Gan, Wentao Zhang, Siguang Li, Shuo Sun, Yunbo Lu, Lin William Cong}}
\makeatletter
\renewcommand\normalsize{\@setfontsize\normalsize{9}{11}%
  \abovedisplayskip 7pt plus 2pt minus 4pt
  \abovedisplayshortskip 0pt plus 2pt
  \belowdisplayshortskip 4pt plus 2pt minus 2pt
  \belowdisplayskip\abovedisplayskip
  \let\@listi\@listI}
\renewcommand\small{\@setfontsize\small{8}{10}}
\renewcommand\footnotesize{\@setfontsize\footnotesize{7}{8}}
\renewcommand\section{\@startsection{section}{1}{\z@}%
  {-8.25pt plus -2pt minus -.2pt}{2.75pt}{\normalfont\fontsize{12}{14}\selectfont\bfseries}}
\renewcommand\subsection{\@startsection{subsection}{2}{\z@}%
  {-8.25pt plus -2pt minus -.2pt}{2.75pt}{\normalfont\fontsize{10}{12}\selectfont\bfseries}}
\makeatother
\normalsize
\renewenvironment{abstract}{\section*{Abstract}\normalsize}{}

\title{\fontsize{18}{21}\selectfont\bfseries M3FT: Micro--Macro--Micro Reasoning for Multi-Agent Cryptocurrency Perpetual Futures Trading}
\author{%
\normalsize Jiashun Wang\textsuperscript{1,*}\quad Beile Wang\textsuperscript{2,*}\quad Guangyan Gan\textsuperscript{3}\quad Wentao Zhang\textsuperscript{3}\\[4pt]
\normalsize Siguang Li\textsuperscript{4}\quad Shuo Sun\textsuperscript{4}\quad Yunbo Lu\textsuperscript{5,\textdagger}\quad Lin William Cong\textsuperscript{3,6,\textdagger}\\[10pt]
\small\textsuperscript{1}Shanghai University of International Business and Economics\\[3pt]
\small\textsuperscript{2}East China Normal University\quad\textsuperscript{3}Nanyang Technological University\\[3pt]
\small\textsuperscript{4}The Hong Kong University of Science and Technology (Guangzhou)\\[3pt]
\small\textsuperscript{5}Tongji University\quad\textsuperscript{6}Tsinghua University\\[6pt]
\footnotesize\textsuperscript{*}These authors contributed equally and share first authorship.\\
\footnotesize\textsuperscript{\textdagger}Corresponding authors.
}
\date{}
\makeatletter
\def\@maketitle{%
  \newpage\null\vskip -8pt
  \begin{center}%
    {\@title\par}\vskip 14pt
    {\normalsize\begin{tabular}[t]{c}\@author\end{tabular}\par}%
  \end{center}%
  \vskip 18.5pt}
\makeatother
\begin{document}
\raggedbottom
\maketitle

\begin{abstract}
Cryptocurrency perpetual futures constitute a large-scale online financial market within the Web economy: in 2025, the world's top ten centralized perpetual futures exchanges recorded \$86.2 trillion in trading volume~\cite{coingecko2025}. Automated trading in this market involves managing multiple positions across assets under changing market conditions to pursue overall account returns. Reinforcement learning learns trading policies from numerical states, while large language models (LLMs) offer a complementary approach through linguistic interpretation of market information. Combining these capabilities presents challenges: policies relying solely on micro-level numerical features may overlook cross-asset relationships and become vulnerable to market regime shifts, while macro-level linguistic interpretation alone cannot ensure precise statistical measurement or optimize positions under trading costs and risk constraints. Reliable trading requires numerical measurement at the micro layer and semantic integration at the macro layer, while enabling LLM agents to accurately interpret numerical states and coordinate their judgments through communication to form joint trading decisions presents a further challenge. To address these challenges, we propose M3FT, a micro--macro--micro framework for multi-agent perpetual futures trading. At the first micro stage, deterministic processing constructs numerical states from exchange and account observations. At the macro stage, LLM agents organize these states into explicit market-level interpretations and exchange asset-specific views, which asset agents refine through communication and review to guide strategy adjustment toward maximizing overall account returns. At the final micro stage, encoded linguistic representations are combined with the original numerical states to learn asset-level trading policies. We implement the framework with a language-conditioned multi-agent policy: each actor uses global and asset-specific numerical states together with encoded linguistic context, while centralized critics evaluate joint actions using portfolio-wide information. Experiments on real cryptocurrency perpetual futures data show that the best-performing M3FT configuration achieves a higher mean test return than the evaluated direct LLM, numerical RL, and rule-based baselines.
\end{abstract}


\section{Introduction}
Cryptocurrency exchanges are digital-native financial platforms within the Web economy, where automated agents participate in continuous trading and allocate capital through online market infrastructure. Cryptocurrency perpetual futures are an important instrument in these markets, enabling leveraged long and short positions without a fixed expiration date. These platforms provide a concrete setting for studying AI-mediated economic decision-making: agents must interpret changing market conditions and translate their judgments into financial actions that affect realized returns and risks. In multi-asset trading, multiple positions interact through shared account capital and margin, so improving individual trading decisions does not necessarily improve overall account returns. Learning how autonomous agents can coordinate these decisions is therefore relevant to capital allocation and risk management in online financial markets.

We define the \emph{micro layer} as numerical states constructed from market and account observations, including asset-specific features and measurable cross-asset relationships. The \emph{macro layer} is an explicit linguistic interpretation of the broader market conditions reflected in these states, rather than an additional source of macroeconomic news. Existing numerical reinforcement learning methods typically learn trading actions directly from numerical observations without such an explicit linguistic intermediary, forming a micro-to-micro decision path. Neither the perpetual-futures portfolio framework of Habibnia and Soltanzadeh~\cite{habibnia2024portfolio} nor FineFT~\cite{qin2025fineft} explicitly incorporates linguistic market interpretation into policy learning. Although these methods can capture numerical dependencies and incorporate risk controls, the interpretation of changing market conditions remains implicit in the learned models. This may make it difficult to transfer learned state--action relationships across market regimes, raising the question of whether explicit market-level interpretation can help policies make more effective use of existing numerical information.

The macro layer cannot replace the micro layer. Language-based trading systems use LLMs to interpret market evidence and synthesize trading judgments. However, QuantAgent~\cite{xiong2025quantagent} has not established how its linguistic market judgments can guide learned joint position optimization across assets. This division of labor between computation and interpretation is important: a language model may produce a plausible account of market conditions, but asking it to inspect raw trading tables and detect statistical changes delegates precise measurement to a generative model. Such analysis may mistake incidental fluctuations for meaningful signals or overlook genuine changes, particularly when inputs are noisy or imperfect. Studies of tabular reasoning have also found that language-model performance can deteriorate in the presence of realistic data artifacts~\cite{gu2025radar}. Moreover, an interpretable judgment alone does not establish how it should inform a learned policy for joint position selection. Reliable trading therefore requires numerical measurement at the micro layer, semantic integration at the macro layer, and a learned mechanism for translating that interpretation back into asset-level actions.

Recent studies have begun to combine LLM reasoning with reinforcement learning, but enabling LLM agents to accurately interpret numerical states and support joint multi-asset position optimization under trading costs and risk constraints remains an additional challenge. Language Model Guided Reinforcement Learning in Quantitative Trading~\cite{darmanin2025language} does not encode a shared linguistic market interpretation for jointly learned multi-asset policies. Adaptive and Explainable Margin Trading via Large Language Models on Portfolio Management~\cite{gu2024margin} uses external market data, leaving the gains from linguistic interpretation insufficiently separated from those of additional inputs. 
\begin{figure}[t]
\centering
\includegraphics[width=\columnwidth]{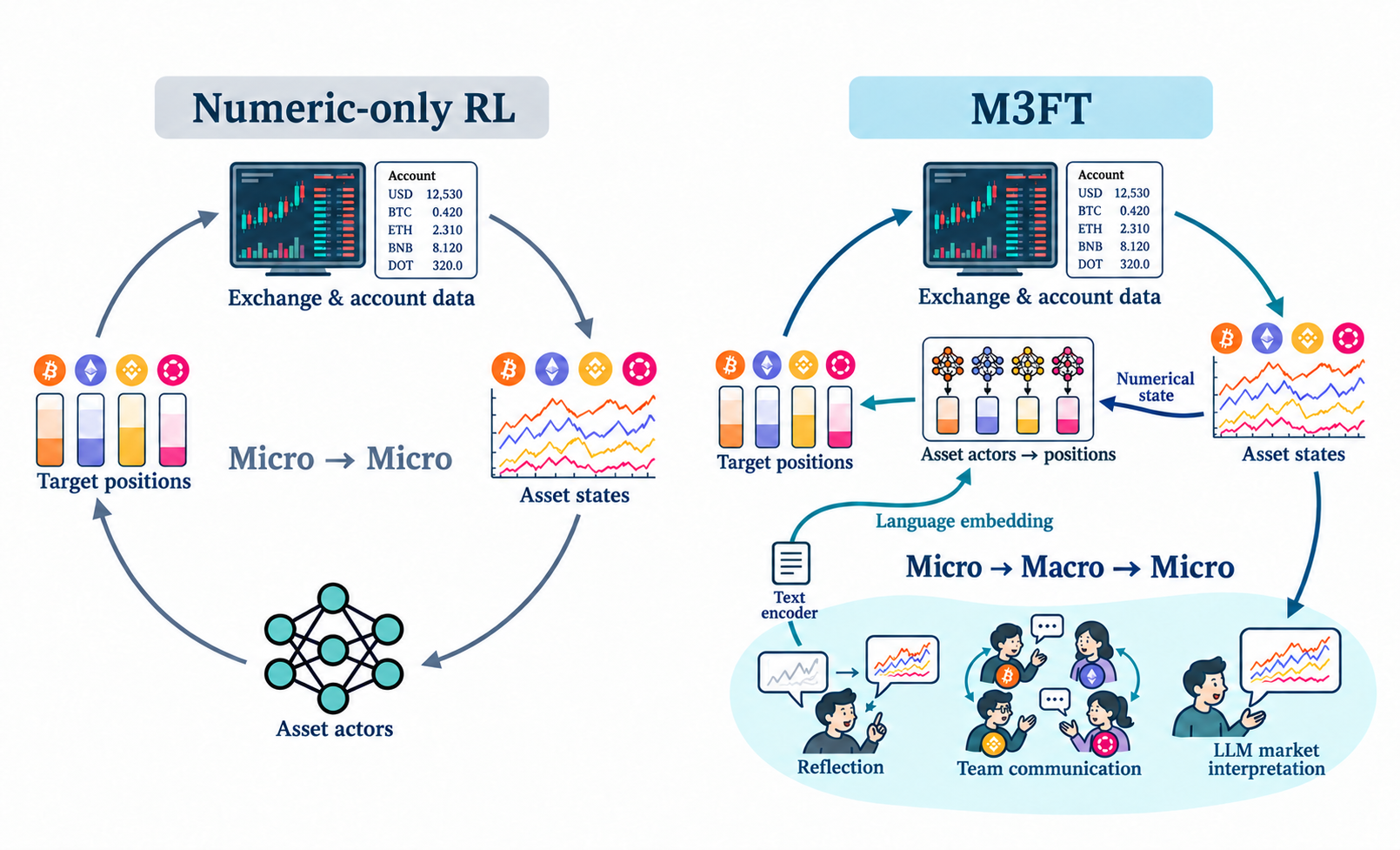}

\caption{Numerical-only RL and M3FT: micro-to-micro versus micro-to-macro-to-micro decision paths.}
\label{fig:intro_micro_macro_micro}
\end{figure}

To address these challenges, we propose \textbf{M3FT}, a micro--macro--micro framework for multi-agent cryptocurrency perpetual futures trading. As illustrated in Figure~\ref{fig:intro_micro_macro_micro}, the framework connects numerical measurement, linguistic market interpretation, and learned position selection. The first micro stage constructs numerical states from exchange and account observations through deterministic processing. The macro stage organizes these measured states into explicit market-level interpretations without introducing additional market observations. At the final micro stage, the resulting language representations are combined with the original numerical states to learn asset-specific trading policies. This design combines grounded market interpretation with numerical policy optimization, allowing us to evaluate whether their integration improves multi-asset trading over numerical-only policies and direct language-model decisions.

To ground LLM interpretations in numerical evidence and support joint position optimization, M3FT separates quantitative measurement, cross-asset review, and policy learning. Deterministic processing computes market and account indicators before language generation, providing the Market Interpreter with measured states and shared market context for asset-specific analysis. The Asset Communicator exchanges position-related information and market views across asset teams. Using these peer assessments together with its own asset's state and exposure, the Risk Reviewer analyzes potential risks arising from joint positions and provides strategy-adjustment guidance toward maximizing overall account returns. The resulting assessments are encoded and combined with numerical states to condition asset-specific actors that select continuous target exposures. Joint actions are evaluated in a shared-account environment that accounts for transaction fees, funding payments, margin requirements, and liquidation risk. Centralized critics use portfolio-wide information and account-level rewards, including trading outcomes and risk penalties, to guide policy updates. This connects grounded linguistic assessments to learned position optimization under trading costs and risk constraints.

Our contributions are threefold:
\begin{enumerate}
    \item We develop M3FT, a language-conditioned micro--\allowbreak macro--\allowbreak micro framework for multi-agent cryptocurrency perpetual-futures trading in a shared cross-margin account. it connects numerical measurement, linguistic market interpretation, and trading policy learning without introducing additional market observations.
    
    \item We design a language-conditioned multi-agent policy that integrates asset-specific numerical states with communication and review, combines peer assessments with each asset’s state and exposure to assess joint-position risks. The resulting guidance is encoded and combined with numerical observations to inform continuous position selection.
    
    \item We validate the framework through extensive experiments on real cryptocurrency perpetual futures data, including comparisons with numerical RL, rule-based, and direct LLM baselines, as well as component ablations.
\end{enumerate}

\section{Related Work}

\subsection{Reinforcement Learning for Multi-Asset Trading}

Reinforcement learning learns sequential decision policies from numerical market states and trading feedback and has been applied to asset allocation and portfolio management. In cryptocurrency futures markets, policies must also account for leverage, transaction costs, and rapidly changing risks. Habibnia and Soltanzadeh~\cite{habibnia2024portfolio} study perpetual-futures portfolio optimization, but do not examine the role of explicit linguistic interpretation in numerical policy learning. FineFT~\cite{qin2025fineft} focuses on training stability and risk management in leveraged futures trading, but does not explore how linguistic interpretation can support cross-asset decisions. We further investigate whether organizing existing numerical observations into market-level semantic context can improve joint multi-asset policy learning within a shared account.

\subsection{LLM-Based Market Interpretation and Trading}

Large language models can organize market information into natural-language judgments, providing explicit analytical grounds for automated trading. Research in this area has expanded from individual trading agents to collaboration among specialized roles and portfolio-level decisions. CryptoTrade~\cite{li2024cryptotrade} explores reflective cryptocurrency trading, but does not incorporate linguistic interpretations into multi-asset reinforcement learning policies. QuantAgent~\cite{xiong2025quantagent} forms trading judgments from price-derived signals, but does not establish the value of these interpretations for joint position policy learning. Luo et al.~\cite{luo2025cryptoportfolio} study multi-agent cryptocurrency portfolio management, but their final trading decisions are not learned through language-conditioned asset-level RL policies. We connect market interpretation with numerical policy learning so that asset-level position choices can be optimized through account-wide feedback.

\subsection{Language-Augmented Reinforcement Learning}

Language-augmented reinforcement learning seeks to combine the semantic analysis capabilities of LLMs with the sequential decision optimization capabilities of RL, connecting market understanding to trading actions. Darmanin and Vella~\cite{darmanin2025language} study LLM-guided reinforcement learning for trading, but their per-instrument guidance does not address joint multi-asset policy learning. Gu et al.~\cite{gu2024margin} combine linguistic market reasoning with reinforcement learning for margin trading, but the introduction of external data leaves the respective benefits of linguistic interpretation and additional market information insufficiently separated. We focus on organizing existing numerical states into linguistic interpretations without adding market observations and using these interpretations for multi-asset policy learning under transaction costs and shared-margin constraints.
\section{Method}

\subsection{Architecture Overview}

To combine precise numerical measurement with market-level interpretation for coordinated trading, we first construct time-aligned numerical states from exchange and account observations. Language agents interpret these measured states and review their implications across asset positions. Their interpretations are then encoded and combined with each asset's original numerical state to learn trading policies, retaining quantitative information for position optimization while incorporating semantic context. This design reorganizes existing observations without introducing additional market inputs. We refer to this information path as the micro--macro--micro framework: numerical states are interpreted in language and then used to learn asset-level trading decisions.

The framework supports different intermediate representations. Our experiments compare representations derived from rules and language using the same exchange and account observations. Below, we describe the language-mediated implementation. Each of the four asset-specific teams contains a \emph{Market Interpreter}, an \emph{Asset Communicator}, and a \emph{Risk Reviewer}. These experts construct language context for asset actors, which learn position decisions through centralized training and decentralized execution.

\begin{figure*}[t]
\centering
\includegraphics[width=\textwidth]{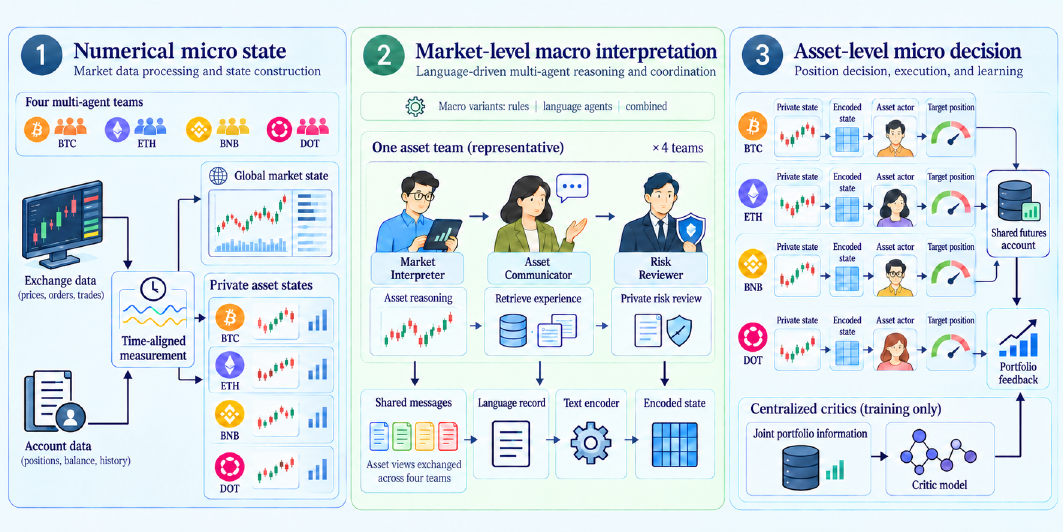}

\caption{Overview of M3FT. Time-aligned exchange and account data provide numerical states for market interpretation and asset-level policies. Encoded language conditions decentralized actors, while centralized critics use joint information during training.}
\label{fig:m3ft_framework}
\end{figure*}

\subsection{Numerical State Construction}

At each decision boundary, a deterministic pipeline aligns heterogeneous exchange and account records and computes complementary numerical states before language generation. The global state summarizes market-wide conditions, cross-contract relationships, and portfolio risk. Each asset team also receives a private state describing its own contract and current exposure. Both the numerical baselines and M3FT make global and asset-specific numerical states available to their decision policies. The language layer interprets these existing inputs without expanding their observable scope.

Each state uses only information available at or before the decision boundary, excluding subsequent prices, funding outcomes, and evaluation results. These numerical observations form the common basis for market interpretation and policy learning. Comparisons between macro representations therefore preserve the underlying exchange and account information while varying how it is organized for decision making.

\subsection{Market Interpretation and Cross-Asset Review}

The macro layer organizes measured market and account conditions into explicit interpretations of their implications for individual assets and joint positions. Its inputs consist of the numerical states, summaries of already observed history, and context previously generated from those observations. No external news, social-media signals, or macroeconomic announcements are supplied to the macro layer. Communication shares interpretations across asset teams, while experience retrieval uses earlier observations and realized feedback from the same episode. Thus, this layer adds a semantic representation of available evidence rather than a new source of market observations.

At a language-refresh event, the Market Interpreters use the common market state, their corresponding asset states, and recent observation summaries. One structured model call produces a shared market description and four asset-specific reasonings. The description integrates market direction, volatility, cross-asset dependence, funding pressure, and portfolio risk; each asset reasoning relates this context to its contract and existing exposure. Refreshes occur at scheduled checkpoints or when monitored volatility, correlation, or aggregate funding changes substantially. On other days, deterministic descriptions grounded in the current observation supply asset-level reasoning without a new language-model call.

Before communication, each team retrieves relevant experiences from the current episode. During training, after a joint action is executed and portfolio feedback becomes available, the system records each asset's numerical-state summary, reasoning, action, private risk review, and realized shared reward. These records enter a capacity-limited language experience pool, which is cleared at each episode boundary and remains separate from the transition replay buffer used for policy updates. At a later refresh, the text encoder embeds current asset reasonings to retrieve similar available records. Earlier interpretations and their observed rewards provide context for the next public statements; retrieved actions are not directly reused as current trading decisions.

The Asset Communicators receive the shared description, preliminary asset reasonings, current numerical summaries, and retrieved experiences. A batched model call generates candidate public statements, and a fixed scoring rule selects one per asset. The four selected statements form a shared message set conveying contract-specific views and their potential cross-asset implications. Communication therefore consists of candidate selection and shared readout, with deterministic selection rather than a learned message-selection network.

The Risk Reviewer connects these exchanged views to the consequences of joint positions in the shared account. Each reviewer examines the positions and market views communicated by the other teams alongside the shared market interpretation, its team's preliminary reasoning, and its own asset's numerical state and exposure. It assesses how the team's trading stance relates to cross-asset dependence, existing portfolio exposure, and shared margin risk. This review places the asset's initial interpretation in the context of the account as a whole and identifies potential risks arising from the combination of positions.

Based on this assessment, the reviewer reconsiders its team's trading stance and provides qualitative guidance for strategy adjustment with the objective of maximizing overall account returns. A final batched model call produces one private review per asset: the teams consider the same peer messages but relate them to their own states and positions. The review conditions the asset actor, which learns how to translate this qualitative guidance into a continuous target exposure using portfolio-level rewards that account for trading outcomes and risk penalties.

For each asset, the preliminary reasoning, selected public statement, and private review form its language record. When no refresh is triggered, the Communicator and Reviewer are skipped, and the current state-grounded reasoning alone forms the record. A fixed text encoder converts each record into a numerical representation, which a trainable conditioning component adapts for the policy.

\subsection{Joint Learning of Asset-Level Trading Policies}

Each asset actor combines its available global and asset-specific numerical states with the conditioned language representation and outputs a continuous target exposure specifying direction and position size. Language experts provide interpretations and adjustment guidance; the actors determine the final positions. Portfolio feedback updates the trading policy and its conditioning component, while the language model and text encoder remain frozen.

Training follows centralized training and decentralized execution. Each asset has its own actor and centralized critic. A critic evaluates its asset's action using the global numerical state, all teams' private and language states, and the joint action, allowing it to learn interactions among positions in the shared account. Each actor selects its position without requiring the centralized critic at execution. All four actors receive the same portfolio-level reward. We instantiate this structure with MADDPG~\cite{lowe2017maddpg} and a MATD3 variant~\cite{ackermann2019matd3}.

The environment evaluates joint target exposures through daily perpetual-futures accounting, including transaction fees, mark-price profit and loss, and funding payments. A conservative liquidation proxy uses adverse within-bar mark prices to represent margin risk. The common reward starts from the change in portfolio equity, which already reflects profit and loss, fees, and funding, with additional penalties for drawdown, excessive leverage, unnecessary turnover, saturated actions, and liquidation. This couples asset-level policy updates to account-wide trading outcomes and risk.

Training proceeds chronologically. Each step constructs numerical states, obtains the available language records, selects joint target exposures, and receives portfolio feedback. Transitions containing numerical and encoded language inputs, joint actions, rewards, and next observations are stored for actor--critic updates. Execution conventions, temporal splits, checkpoint selection, and evaluation settings are specified in the experimental protocol.

\begingroup
\setlength{\emergencystretch}{3em}
\section{Experiments}
\label{sec:experiments}

\subsection{Research Questions}
Our experiments examine whether an explicit intermediate market-level
abstraction improves cryptocurrency perpetual-futures trading. We organize the
evaluation around four research questions.

\noindent\textbf{RQ1: Does the micro--macro--micro architecture improve trading
performance over direct micro-to-action policies?} We compare M3FT with
numerical-only reinforcement-learning policies that map numerical observations
directly to actions.

\noindent\textbf{RQ2: Is an intermediate learned language representation more
effective than direct LLM trading and deterministic rule-based abstraction?}
We compare M3FT with LLMs that directly output positions and with two
rule-based macro-state constructions.

\noindent\textbf{RQ3: Are the benefits of M3FT preserved across policy-learning
algorithms and language-model providers?} We instantiate the framework with
MADDPG and MATD3 and evaluate both DeepSeek and GPT language models.

\noindent\textbf{RQ4: Which language-layer components account for the observed
behavior?} We remove the Asset Communicator and Risk Reviewer together in one
controlled ablation and remove the Experience Pool in another. 

\subsection{Dataset and Trading Environment}
We use daily Binance USD{\fontfamily{cmr}\selectfont\textcircled{S}}-M perpetual-futures records for
BTCUSDT, ETHUSDT, BNBUSDT, and DOTUSDT~\cite{binance2026data}. After temporal alignment, the dataset
contains 2,008 common trading dates from January 31, 2021 to July 31, 2026.
The temporal cut points are May 18, 2024 and June 24, 2025, producing
non-overlapping training, validation, and held-out test periods. The data are
never randomly shuffled across these boundaries.

All methods use the same time-aligned exchange and account records. Numerical
baselines and M3FT have access to the same global and asset-specific numerical
states, including cross-asset relationships and shared-account conditions.
These inputs are available to the numerical decision policies, not only to
the centralized critics. The observations include price-derived features, volatility, volume
change, taker-buy activity, funding, premium, cross-asset dependence, and the
current account state. No method receives future prices, future funding
outcomes, test labels, external news, social-media signals, or macroeconomic
announcements. M3FT differs in representation rather than information
availability: it converts deterministic summaries of the same observed records
into an additional language-derived representation.

The four contracts are traded in one shared cross-margin account. Each decision
is made using information available at the end of a UTC day. The first recorded
price in the following daily bar is used as the execution-price proxy. The
environment then accounts for transaction fees, mark-price profit and loss,
funding payments, portfolio equity, margin usage, and liquidation risk. Each
learned actor outputs either a continuous target exposure or, for the adapted
CPTS baseline, one of four discrete position instructions. Continuous actions
lie in $[-1,1]$, where the sign determines direction and the magnitude controls
the target exposure.

We initialize each episode with an account equity of $1{,}000{,}000$, use a
threefold target-notional budget at rebalancing, and charge a transaction-fee
rate of $0.0004$. All
learned methods use the same environment, reward accounting, temporal
boundaries, and held-out test protocol.

\subsection{Controlled Training and Evaluation Protocol}
Formal experiments use four independent random seeds, denoted A--D in the
paper. MADDPG- and MATD3-based configurations, PPO, and the adapted CPTS core
are trained for at most ten epochs. Each epoch contains eight training episodes
of 120 daily steps and evaluation on two fixed validation windows. Training
stops after three consecutive epochs without improvement. Checkpoint selection
depends exclusively on validation performance and never on the test results.

After model selection, the selected checkpoint is evaluated without parameter
updates or exploration noise. All methods use the same six fixed 120-step test episodes
per seed. We first average the six episode-level results within each seed and
then report the mean and sample standard deviation across the four seeds.
The random seed is the statistical unit; the six test episodes are not treated
as six independent seeds.

The LLM itself and the text encoder remain frozen during reinforcement
learning. Portfolio rewards update the trading policy and its
language-conditioning module, but do not fine-tune the provider model. Model
selection, early stopping, prompts, environment parameters, data boundaries,
and test windows remain fixed within each controlled comparison.

\subsection{Compared Methods}
We divide the compared methods into three levels according to the path from
market observations to trading actions.

\paragraph{M1: Direct LLM decision making.}
A language model observes the available market and account summary and directly
produces the four target actions in one structured request per decision step.
This baseline has no reinforcement-learning policy, training phase, validation
selection, text embedding, Market Interpreter, Asset Communicator, Risk
Reviewer, or Experience Pool. We evaluate both DeepSeek Flash~\cite{deepseek2026flash} and
GPT-5.6-terra~\cite{openai2026terra}. M1 tests whether language reasoning alone can translate
observations into executable positions.

\paragraph{M2: Numerical policy learning.}
These methods map numerical observations to actions without an explicit
intermediate macro layer.

\emph{PPO}~\cite{schulman2017ppo} uses a multilayer perceptron policy with a 42-di\-men\-sion\-al flattened
numerical observation and four continuous target-position outputs. We use $\gamma=0.99$, 120 rollout steps, a batch size of
60, ten PPO optimization epochs, and two hidden layers of 128 units.

\emph{MADDPG}~\cite{lowe2017maddpg} uses one continuous actor per contract and centralized critics
for portfolio-level learning. We use $\gamma=0.99$ and $\tau=0.005$.

\emph{MATD3}~\cite{ackermann2019matd3} extends the multi-agent deterministic policy with twin critics,
delayed policy updates, and smoothed target actions~\cite{fujimoto2018td3}. We use policy delay 2, target noise 0.2 clipped to 0.5, Huber loss~\cite{huber1964robust}
with $\beta=1$, and gradient clipping at 1.

\emph{CPTS-adapted A2C-LSTM} retains the core structure described by Chun and
Lee~\cite{chun2025cpts}: independent asset-level A2C policies~\cite{wu2017a2c},
two-layer LSTM actors and critics~\cite{hochreiter1997lstm},
and the discrete actions enter-long, enter-short, hold, and close. It uses
$\gamma=0.99$, entropy coefficient 0.01, and hidden
size 64. For controlled comparison, it is adapted to the same four daily
contracts, shared futures account, reward, and temporal split. It does not
reproduce the original system's 18-asset selection, multiple frequencies,
ANOVA stage, or additional on-chain variables; we therefore refer to it as an
adapted CPTS core rather than a full CPTS reproduction.

\paragraph{M3: Micro--macro--micro decision making.}
M3 methods insert a market-level representation between numerical measurement
and asset-level action selection.

The two rule-based controls abstract the language module's market judgments
into deterministic numerical rules. They receive the same numerical market and
account inputs as the other learned methods and pass a rule-derived macro
interpretation to the same RL execution layer. This design tests whether the
observed gains arise from the way the language layer organizes measured
information, rather than merely from giving the policy an additional macro
state.

\emph{Rule-state + MADDPG} constructs a nine-dimensional deterministic macro
state. Market variables are discretized into low, neutral, and high states
using quartiles estimated from the training period. Its channels describe
return breadth, volatility, correlation, volume change, taker-buy activity,
funding, premium change, drawdown risk, and margin-usage risk.

\emph{Bollinger-state + MADDPG} constructs a deterministic macro state from a
20-day rolling mean and two population standard deviations~\cite{bollinger2001bands}. It provides a
clipped price-deviation score and a contrarian band indicator, while unused
channels are set to zero to preserve the same nine-dimensional interface.
Signals are formed at day $t$ and acted on at the following execution boundary.

\emph{M3FT} replaces the fixed rule abstraction with a language-mediated market
representation. A Market Interpreter summarizes market-wide and asset-specific
conditions; Asset Communicators exchange selected contract-level views; Risk
Reviewers reconsider each asset's interpretation in light of the shared
messages; and an Experience Pool retrieves relevant cases from the current
episode. A frozen \texttt{text-embedding-v4} encoder~\cite{alibaba2026embedding} maps each asset-specific language
record to a 64-dimensional representation. The actor combines this representation with the available global and
asset-specific numerical states and remains solely responsible for the final
position.

The compared M3FT variants cross two language models, DeepSeek and GPT, with
MADDPG and MATD3. Thus, the evaluation tests the framework across both provider
choice and reinforcement-learning dynamics.

\subsection{Evaluation Metrics}
\label{sec:metrics}
Let $E_t$ denote portfolio equity after accounting for trading profit and loss,
fees, and funding, and let $r_t=\log(E_t/E_{t-1})$ denote the daily log return.

We report total test return,
\begin{equation}
R=\frac{E_T}{E_0}-1,
\end{equation}
and the Sharpe ratio under a 252-day scaling convention~\cite{sharpe1994ratio},
\begin{equation}
\operatorname{Sharpe}=\sqrt{252}\frac{\bar r}{s_r},
\end{equation}
where $\bar r$ and $s_r$ are the mean and population standard deviation
of daily log returns, respectively. The risk-free rate is set to zero.

Maximum drawdown is reported as a positive loss magnitude,
\begin{equation}
\operatorname{MDD}=-\min_t\left(\frac{E_t}{\max_{u\leq t}E_u}-1\right).
\end{equation}

Action saturation measures the fraction of decisions placed in the
near-boundary region of the action space. Because continuous deterministic
policies and the PPO/CPTS evaluators use slightly different native boundary
conventions, saturation is interpreted as a policy-behavior diagnostic rather
than as a primary cross-algorithm performance measure.

Average gross leverage is
\begin{equation}
\overline{L}_{\mathrm{gross}}=
\frac{1}{T}\sum_{t=1}^{T}\frac{\sum_i |V_{i,t}|}{E_t},
\end{equation}
Average absolute net exposure, normalized by the leverage budget, is
\begin{equation}
\overline{L}_{\mathrm{net}}=
\frac{1}{T}\sum_{t=1}^{T}
\left|\frac{\sum_i V_{i,t}}{\lambda E_t}\right|,
\end{equation}
where $\lambda=3$ is the configured leverage multiplier,
so $\lambda E_t$ denotes the leverage budget at step $t$.

We additionally inspect liquidation, non-finite values, and loss instability.
For $K=4$ seeds, reported dispersion is the sample standard deviation,
\begin{equation}
s_x=\sqrt{\frac{1}{K-1}\sum_{k=1}^{K}(x_k-\bar{x})^2}.
\end{equation}

\begin{table*}[!t]
\centering
\captionsetup{position=top}
\caption{Out-of-sample comparison of trading performance, risk, and position behavior across direct LLM, numerical RL, rule-based, and M3FT policies.}
\label{tab:main_results}
{\footnotesize
\setlength{\tabcolsep}{3pt}
\renewcommand{\arraystretch}{1.18}
\begin{tabular*}{0.90\textwidth}{@{\extracolsep{\fill}}lcccccc}
\toprule
& \multicolumn{2}{c}{Performance} & \multicolumn{1}{c}{Risk} & \multicolumn{3}{c}{Position behavior} \\
\cmidrule(lr){2-3}\cmidrule(lr){4-4}\cmidrule(l){5-7}
Method & Test return (\%)$\uparrow$ & Sharpe$\uparrow$ & Drawdown (\%)$\downarrow$ & Sat. (\%) & Gross lev. & $|$Net exp.$|$ \\
\midrule
\multicolumn{7}{l}{\textit{M1: Direct LLM trading}} \\
GPT-5.6-terra & $+4.55\pm3.26$ & $0.34\pm0.22$ & $23.96\pm2.14$ & $0.00\pm0.00$ & $0.65\pm0.01$ & $0.18\pm0.00$ \\
DeepSeek Flash & $+4.55\pm16.61$ & $0.23\pm1.06$ & $24.19\pm4.58$ & $0.36\pm0.26$ & $0.65\pm0.03$ & $0.18\pm0.01$ \\
\midrule
\multicolumn{7}{l}{\textit{M2: Numerical policy learning}} \\
MATD3 & $-9.06\pm13.11$ & $-0.67\pm0.85$ & $24.12\pm8.28$ & $0.89\pm1.77$ & $1.03\pm0.32$ & $0.16\pm0.05$ \\
MADDPG & $-7.49\pm3.88$ & $-0.80\pm0.47$ & $26.25\pm9.89$ & $4.43\pm7.12$ & $1.15\pm0.36$ & $0.22\pm0.14$ \\
PPO & $-0.85\pm2.85$ & $-0.61\pm0.61$ & $16.53\pm8.13$ & $2.73\pm5.33$ & $0.83\pm0.58$ & $0.08\pm0.07$ \\
CPTS-adapted A2C-LSTM & $-0.89\pm15.85$ & $0.20\pm1.18$ & $23.84\pm20.81$ & $30.49\pm25.45$ & $0.92\pm0.77$ & $0.18\pm0.25$ \\
\midrule
\multicolumn{7}{l}{\textit{M3: Rule-based macro abstraction}} \\
Bollinger-state + MADDPG & $-9.09\pm16.00$ & $-0.74\pm0.88$ & $34.82\pm5.27$ & $8.28\pm10.92$ & $1.30\pm0.21$ & $0.24\pm0.09$ \\
Rule-state + MADDPG & $-2.41\pm14.61$ & $-0.18\pm0.84$ & $32.68\pm9.65$ & $4.56\pm1.92$ & $1.38\pm0.10$ & $0.22\pm0.04$ \\
\midrule
\multicolumn{7}{l}{\textit{M3: Language-mediated M3FT}} \\
DeepSeek + MATD3 & $+4.25\pm10.15$ & $-0.12\pm0.92$ & $29.27\pm16.68$ & $6.93\pm12.12$ & $1.39\pm0.45$ & $0.20\pm0.15$ \\
GPT + MATD3 & $+5.64\pm10.87$ & $0.34\pm0.87$ & $19.40\pm10.84$ & $1.88\pm2.03$ & $1.083\pm0.22$ & $0.18\pm0.14$ \\
\textbf{DeepSeek + MADDPG} & $+9.87\pm16.22$ & $0.37\pm0.56$ & $29.20\pm3.86$ & $81.15\pm12.57$ & $2.68\pm0.27$ & $0.11\pm0.09$ \\
GPT + MADDPG & $+6.21\pm18.45$ & $0.12\pm0.82$ & $30.49\pm8.41$ & $76.75\pm25.18$ & $2.48\pm0.69$ & $0.19\pm0.26$ \\
\bottomrule
\end{tabular*}
}\par
\end{table*}

\section{Experiment Results}
\label{sec:experiment_results}
\subsection{Main Results}
\label{sec:main_results}
Table~\ref{tab:main_results} reports the principal results as mean $\pm$
sample standard deviation across four independent random seeds.

The clearest pattern is the separation between numerical-only M2 policies and
language-mediated M3FT policies. All four M2 policies have negative mean test
returns: MATD3 obtains $-9.06\%$, MADDPG $-7.49\%$, PPO $-0.85\%$, and the
adapted CPTS core $-0.89\%$. By contrast, all four language-mediated M3FT
configurations obtain positive mean returns: DeepSeek\allowbreak{}+\allowbreak{}MATD3 reaches $4.25\%$,
GPT\allowbreak{}+\allowbreak{}MATD3 $5.64\%$, DeepSeek\allowbreak{}+\allowbreak{}MADDPG $9.87\%$, and GPT\allowbreak{}+\allowbreak{}MADDPG $6.21\%$.

Relative to numerical-only MADDPG, the DeepSeek and GPT M3FT variants improve
mean Return by 17.36 and 13.70 percentage points, respectively. Relative to
numerical-only MATD3, the corresponding improvements are 13.31 and 14.70
percentage points. Because the numerical observation, environment, reward,
and evaluation period are controlled, this repeated direction of improvement
supports RQ1: explicitly passing measured micro states through a market-level
language representation can improve the subsequent asset-level policy.

The M1 direct-LLM baselines both achieve a mean Return of $4.55\%$, showing
that language models can produce useful actions even without policy learning.
Their behavior, however, does not establish that direct LLM control is
uniformly preferable. Three of the four M3FT configurations exceed the
direct-LLM mean Return. GPT\allowbreak{}+\allowbreak{}MATD3 improves Return to $5.64\%$, maintains a
Sharpe ratio of $0.34$, and reduces MDD from $23.96\%$ for direct GPT to
$19.40\%$. DeepSeek\allowbreak{}+\allowbreak{}MADDPG raises Return to $9.87\%$, although this gain is
accompanied by greater leverage and action saturation. DeepSeek\allowbreak{}+\allowbreak{}MATD3 is
slightly below direct DeepSeek in mean Return, demonstrating that the benefit
of language-conditioned policy learning is not identical for every
provider--learner pairing.

The deterministic M3 rule baselines do not produce the same improvement.
Bollinger-state + MADDPG records a mean Return of $-9.09\%$, while Rule-state
+ MADDPG improves this to $-2.41\%$ but remains negative. The richer
Rule-state representation is therefore more effective than the narrow
Bollinger abstraction, yet both remain below the language-mediated variants.
This answers RQ2: inserting any intermediate representation is insufficient by
itself. The principal benefit appears when the macro layer can integrate
several measured market and account conditions into an asset-conditioned
representation rather than compressing them into a small fixed set of
thresholds.

The comparison also reveals a return--risk trade-off. DeepSeek\allowbreak{}+\allowbreak{}MADDPG delivers
the highest mean Return, but its saturation rate of $81.15\%$ and gross
leverage of 2.68 indicate an aggressive policy. GPT\allowbreak{}+\allowbreak{}MADDPG shows a similar
pattern, with $76.75\%$ saturation and gross leverage of 2.48. MATD3 produces
more restrained policies: GPT\allowbreak{}+\allowbreak{}MATD3 combines a positive $5.64\%$ Return with
Sharpe $0.34$, MDD $19.40\%$, saturation $1.88\%$, and gross leverage 1.083.
Accordingly, MADDPG provides the strongest raw-return result, whereas
GPT\allowbreak{}+\allowbreak{}MATD3 gives the most balanced language-mediated outcome in the reported
risk diagnostics.

For RQ3, all four M3FT configurations achieve positive mean returns and
outperform their numerical-only counterparts in mean return across both
language providers and policy learners. However, the large cross-seed
standard deviations limit conclusions about consistency across seeds,
and no configuration dominates every risk metric.

\begin{table*}[!t]
\centering
\captionsetup{position=top}
\caption{Effects of removing communication and review or experience retrieval on trading performance, risk, and position behavior in DeepSeek+MADDPG.}
\label{tab:ablations}
{\footnotesize
\setlength{\tabcolsep}{3pt}
\renewcommand{\arraystretch}{1.18}
\begin{tabular*}{\textwidth}{@{\extracolsep{\fill}}p{1.65in}cccccc}
\toprule
Variant & Return (\%)$\uparrow$ & Sharpe$\uparrow$ & MDD (\%)$\downarrow$ & Sat. (\%) & Gross lev. & $|$Net exp.$|$ \\
\midrule
Full model & $+9.87\pm16.22$ & $0.37\pm0.56$ & $29.20\pm3.86$ & $81.15\pm12.57$ & $2.68\pm0.27$ & $0.11\pm0.09$ \\
Without Asset Communicator and Risk Reviewer & $+7.02\pm25.07$ & $-0.112\pm0.690$ & $39.55\pm14.99$ & $100.00$ & $3.006$ & $0.337$ \\
Without Experience Pool & $-6.27\pm2.51$ & $-0.379\pm0.205$ & $34.34\pm14.24$ & $100.00$ & $3.006$ & $0.131$ \\
\bottomrule
\end{tabular*}
}\par
\end{table*}

\begin{figure}[t]
    \centering
    \includegraphics[width=\columnwidth]{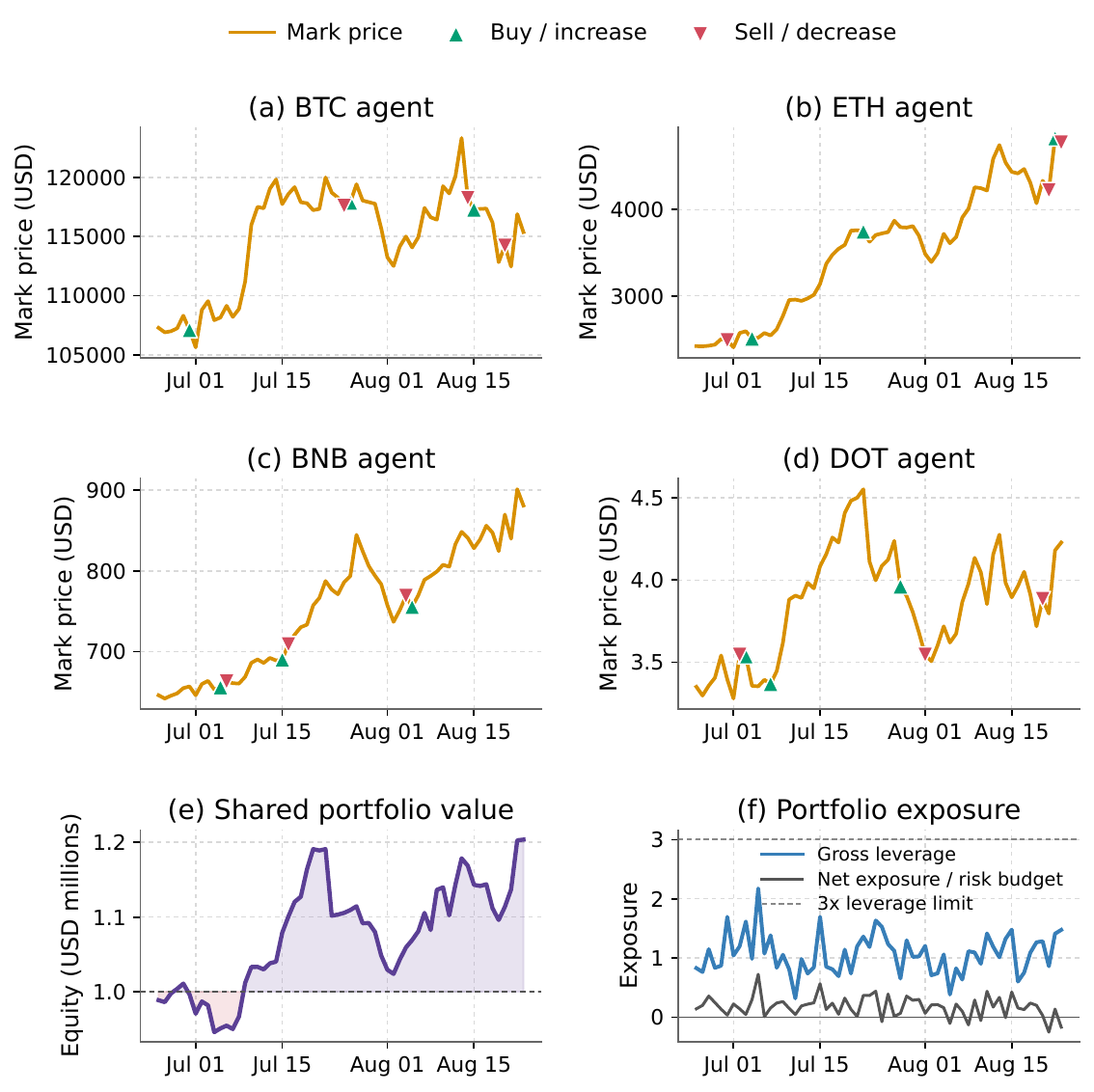}
    \caption{A representative four-agent M3FT trading episode, showing
    asset-level target-exposure changes and shared-account portfolio dynamics.}
    
    \label{fig:m3ft_trading_case}
\end{figure}

\subsection{Component Ablations}
\label{sec:ablations}
Table~\ref{tab:ablations} compares the complete DeepSeek+MADDPG configuration
with two component removals using four seeds. The full-model row uses exactly
the same reference results as Table~\ref{tab:main_results}. The comparison
jointly removes the Asset Communicator and Risk Reviewer in one variant and
removes the Experience Pool in the other. The full model and both ablation
variants use the same training and evaluation configuration, including
\texttt{text-embedding-v4} and the checkpoint-selection rule; only the
specified components are removed. Return, Sharpe, and MDD are reported as
mean $\pm$ sample standard deviation across four seeds. Position diagnostics
include standard deviations for the full model and means for the
component-removal variants.

Removing the Experience Pool turns mean Return negative, lowers Sharpe,
and increases drawdown and action saturation. These results support the
role of experience retrieval in the language layer: referring to agents'
previous inferences and their realized outcomes provides a basis for
comparison when forming current market judgments, supporting more effective
trading and reducing extreme position behavior.

Removing the Asset Communicator and Risk Reviewer lowers mean Return to
$7.02\%$ and Sharpe to $-0.112$, while increasing MDD to $39.55\%$.
These results indicate that retaining language-based interpretations of
numerical market states can still yield positive mean returns, while
information exchange among asset positions and review of cross-asset
exposures further improve shared-account trading performance and stability,
with less extreme position behavior.

These ablation results answer RQ4 by showing how experience retrieval and
cross-asset communication and review support the effectiveness of the
language layer. Its contribution extends beyond interpreting current
numerical market states: it also draws on past
experience and uses cross-asset communication and review to place
asset-specific judgments within the broader context of the shared account.
These mechanisms jointly support more effective and stable multi-asset
position decisions.

\subsection{Representative M3FT Trading Behavior}
\label{sec:trading_case}

Figure~\ref{fig:m3ft_trading_case} illustrates one representative fixed test
episode from the GPT+MATD3 M3FT configuration. Its purpose is to make the
M3FT micro--macro--micro decision process concrete rather than to support
a cross-method performance claim. The top four panels show BTC, ETH, BNB,
and DOT mark prices over a 60-day segment. Each of the four asset actors
outputs a continuous target exposure; green (red) triangles mark the three
largest daily increases (decreases) in each actor's target exposure within
this segment. Decisions formed at the day-$t$ close are executed at the
day-$(t+1)$ open proxy.

The actors rebalance at different times, but their
positions are settled through one shared cross-margin account; the lower
panels therefore report joint portfolio equity, gross leverage, and net
exposure. Unlike the numerical-only MATD3 baseline, M3FT additionally
conditions each actor on an encoded market-level interpretation of the
same measured inputs.The asynchronous rebalancing across panels further clarifies the role of the macro layer. A shared market interpretation coordinates the four actors without forcing identical actions: each actor combines the common macro embedding with its own contract-level state, thereby preserving asset-specific responses within a portfolio-level view.

\subsection{Efficiency, Limitations, and Scope of Interpretation}
M3FT incurs more inference time and provider cost than numerical-only policies.
The system mitigates this overhead through event-triggered language refresh and
batched requests across the four assets. A new language pass is generated at
scheduled checkpoints or when monitored volatility, correlation, or funding
conditions change materially. On ordinary steps, the policy reuses the
available context and deterministic state-grounded descriptions; it does not
invoke the LLM at every market step. This mechanism reduces remote calls while
preserving the macro representation during relatively stable periods.

Several additional limitations define the scope of the conclusions. The
evaluation covers four liquid perpetual contracts at daily frequency and uses
a historical simulator rather than live exchange execution. The next-bar
execution proxy cannot reproduce order-book depth, latency, slippage, or
partial fills. Language-model outputs may vary across providers and service
versions even under fixed prompts. The large return dispersion and the
aggressive behavior of some MADDPG variants further show that the macro layer
does not remove market and optimization risk. Future work should evaluate more
contracts, higher-frequency observations, execution-aware simulation,
additional market periods, and explicit constraints that preserve the
representational benefit of M3FT while limiting excessive leverage and
saturation.

Future work will explore more token-efficient market summaries
and cross-asset communication to reduce inference costs while
preserving decision-relevant information and trading performance.
\endgroup



\clearpage
\bibliographystyle{ACM-Reference-Format}
\bibliography{references}

\end{document}